\documentclass[prd,twocolumn,preprintnumbers,floatfix,superscriptaddress]{revtex4}
\usepackage[colorlinks=true,linkcolor=red,citecolor=blue,urlcolor=darkred]{hyperref}
\usepackage{gensymb,comment,cancel}
\usepackage{amsmath,amssymb,epsfig,bm,mathrsfs,color}
\usepackage{slashed}

\newcommand{\be}{\begin{equation}}
\newcommand{\ee}{\end{equation}}
\newcommand{\bea}{\begin{eqnarray}}
\newcommand{\eea}{\end{eqnarray}}

\newcommand{\ie}{{\it i.e.}}

\newcommand{\MeV}{\text{MeV}}
\newcommand{\kHz}{\text{kHz}}

\newcommand{\mnras}{Mon.~Not.~RAS}

\definecolor{red}{rgb}{0.8,0,0}
\definecolor{orange}{rgb}{0.8,0.2,0.0}
\definecolor{blue}{rgb}{0.3,0.0,0.8}
\definecolor{green}{rgb}{0,0.5,0.0}
\definecolor{darkred}{rgb}{0.7,.1,.2}
\definecolor{bgred}{rgb}{1.,.95,.95}
\definecolor{bgblue}{rgb}{.95,.95,1.}
\definecolor{bluegreen}{rgb}{0.,.5,.3}
\definecolor{darkred}{rgb}{0.7,.1,.2}
\definecolor{darkgreen}{rgb}{0.1,.6,.0}
\definecolor{lightyellow}{rgb}{1.,1.,.8}
\definecolor{darkcyan}{rgb}{0.,.7,.9}
\definecolor{lightblue}{rgb}{0.6,0.8,1}
\definecolor{lightgreen}{rgb}{0.7,1.,.9}
\definecolor{money}{rgb}{0.4,0.8,0.}
\definecolor{purple}{rgb}{0.9,0.0,0.8}
\definecolor{orange}{rgb}{0.9,0.5,0.0}
\definecolor{newgr}{rgb}{0.2,0.8,0.2}
\definecolor{newbl}{rgb}{0.3,0.6,0.8}
\definecolor{newor}{rgb}{1.0,0.6,0.}

\usepackage[normalem]{ulem}  % \sout{old text} for strikeout  

\begin{document}

\title{ Bulk viscosity of two-color superconducting quark matter in neutron star mergers}

\author{Mark Alford} \email{alford@physics.wustl.edu}
\affiliation{Department of Physics, Washington University, St.~Louis,
  Missouri 63130, USA}

\author{ Arus Harutyunyan} \email{arus@bao.sci.am}
\affiliation{Byurakan Astrophysical Observatory,
  Byurakan 0213, Armenia}
\affiliation{Department of Physics, Yerevan State University, Yerevan 0025, Armenia}

\author{ Armen Sedrakian}\email{sedrakian@fias.uni-frankfurt.de}

\affiliation{Institute of Theoretical Physics, University of Wroc\l{}aw,
50-204 Wroc\l{}aw, Poland}
\affiliation{Frankfurt Institute for Advanced Studies, D-60438
  Frankfurt am Main, Germany}

\author{Stefanos Tsiopelas} \email{stefanos.tsiopelas2@uwr.edu.pl}
\affiliation{Institute of Theoretical Physics, University of Wroc\l{}aw,
50-204 Wroc\l{}aw, Poland}

\date{22 August 2024} 

\begin{abstract}

  We study the bulk viscosity of moderately hot and dense,
  neutrino-transparent color superconducting quark matter arising from
  weak-interaction-driven direct URCA processes. The quark matter is
  modeled using the Nambu--Jona-Lasinio model improved to account for
  vector and 't Hooft interactions as well as antisymmetric pairing
  among the red/green up and down quarks.  The unpaired excitations
  are the strange quarks and the blue up or down quarks. We compute
  the relaxation rates associated with $d$ and $s$-quark decay and
  electron capture processes on $u$ quark for blue color. The
  resulting bulk viscosity for density oscillations in the 1--10\,kHz
  range shows a resonant peak at $T\sim 5\,\MeV$, and the damping time
  may drop below 10 ms.  This is short enough to affect the
  postmerger evolution and is very similar to the damping predicted
  in nuclear matter.
\end{abstract}
%\keyword{URCA processes; bulk viscosity; neutrino-trapping; neutron star mergers}
%\pacs{21.65.+f, 21.30.Fe, 26.60.+c}

\maketitle

\section{Introduction}
\label{sec:intro}

The recent discovery of gravitational waves followed by
electromagnetic counterparts originating from binary neutron star
mergers, as observed by the LIGO-Virgo
Collaboration~\cite{Abbott2017}, has spurred interest in the transport
characteristics of hot and dense matter. Numerical simulations of
binary neutron star mergers typically conducted within the framework
of nondissipative hydrodynamics have predicted significant density
oscillations and robust gravitational wave emissions in the initial
tens of milliseconds following the
merger~\cite{Faber2012,Baiotti2017,Baiotti2019,Bauswein2019,Radice2023,Radice2024}. These
oscillations are expected to eventually dissipate due to various
dissipative processes within the postmerger matter, ultimately
impacting the gravitational wave signal. Among these processes, bulk
viscous dissipation facilitated by weak interactions appears to be the
most efficacious mechanism in damping these oscillations as
demonstrated by the initial estimates~\cite{Alford2018a},
first-principle computations of rates and associated relaxations
times~\cite{Alford2020,Alford2021a,Alford2021c,Alford2023} and
dissipative numerical
simulations~\cite{Most2022,Celora2022,Chabanov2023}.

In the cold regime pertinent to mature compact stars, extensive
studies on bulk viscosity have been conducted, largely building upon
the seminal work of Ref. \cite{Sawyer1989}. The case of cold quark
matter in hybrid and strange stars was studied in
Refs.~\cite{Madsen1992,Drago2005,Alford:2006gy,Blaschke2007,Sad2007a,Sad2007b,Huang2010,Wang2010a,Wang2010b}
in the context of damping of $r$-mode oscillations of cold stars. In
particular, two-flavor (``2SC'') pairing was studied
in~Ref.~\cite{Alford:2006gy} including finite temperatures within a
model defined in terms of the 2SC gap and chemical potentials (without
electrical neutrality) while the numerical results were focused on
non-leptonic processes \footnote{See also Appendix A of
Ref.~\cite{Alford:2006gy} where the bulk viscosity is derived for
coupled nonleptonic and leptonic processes, similar to our approach
here.}.

The study of bulk viscosity of quark matter in the context of binary
neutron star (BNS) mergers is still in its early stages
\cite{Rojas2024, Hernandez2024} and has focused on hot, unpaired quark
matter. In this work, we present the first computation of the bulk
viscosity of a BNS postmerger object that incorporates color
superconductivity at intermediate densities and nonzero
temperatures. We also consider conditions specific to BNS mergers,
such as charge and color neutrality, and the density and temperature
dependence of gaps and chemical potentials, which are accounted for
self-consistently within our model.

We focus on the weak leptonic quark URCA processes within the phase
structure of color superconducting matter relevant to compact
stars. At intermediate densities, quark matter is expected to be in
the two-color superconducting (2SC) phase, where red-green light
quarks pair while light blue quarks, as well as $s$ quarks, remain
unpaired~\cite{Alford2007}. Our study, using the
vector-interaction-improved Nambu--Jona-Lasinio (NJL)
model~\cite{Bonanno2012}, examines bulk viscosity primarily from the
URCA processes involving these unpaired blue quarks.

\section{Bulk viscosity of $udse$ matter}
\label{sec:bulk}

We consider neutrino-transparent matter composed of $u, d, s$ quarks
and electrons in the core of the compact star. The simplest
(semileptonic) $\beta$ equilibration processes in this system are the $d$
and $s$-quark decay and the electron capture processes of the direct
URCA type,
%------------------------------------------------------
\bea\label{eq:d_decay}
&& d\rightarrow u + e^-+\bar{\nu}_e,\\
\label{eq:e_capture_d}
&& u + e^-\rightarrow d +{\nu}_e,\\
\label{eq:s_decay}
&& s\rightarrow u + e^-+\bar{\nu}_e,\\
\label{eq:e_capture_s}
&& u + e^-\rightarrow s +{\nu}_e,
\eea
%------------------------------------------------------
where $\nu_e$ and $\bar\nu_e$ are the electron neutrino and antineutrino, respectively. The processes~\eqref{eq:d_decay}--\eqref{eq:e_capture_s}
proceed only in the direction from left to right because in 
neutrino-transparent matter neutrinos/antineutrinos can appear only in final states. 

The rates of the URCA processes \eqref{eq:d_decay}--\eqref{eq:e_capture_s} can be written in full analogy with the nucleonic counterparts, see for example, Eqs.~(8)--(10) of~\cite{Alford2021c} (for quark matter one simply needs to replace $g_A\to 1$ in the matrix element and also $\cos\theta_c\to \sin\theta_c$ for processes including $s$-quarks, where $\theta_c=13.04^o$ is the Cabibbo angle).

We will show below that, as has previously been noted \cite{Alford:2006gy,Rojas2024,Hernandez2024}, the nonleptonic processes 
%------------------------------------------------------
\bea\label{eq:non-leptonic}
u+d\leftrightarrow u+s
\eea
%------------------------------------------------------
 proceed much faster than the URCA processes. Consequently, in the regime where URCA-process-driven bulk viscosity is at its maximum the non-leptonic equilibration effectively takes place immediately, and we can assume  $\mu_s=\mu_d$ at all times (see~\cite{Alford:2006gy}, Appendix A).
In this regime, there is only a single equilibrating quantity
%------------------------------------------------------
%\bea\label{eq:mu_Delta}
$\mu_\Delta\equiv\mu_d-\mu_u-\mu_e$,
%\eea
%------------------------------------------------------
which is the relevant measure of how much the system is driven out of $\beta$-equilibrium state by a cycle of compression and rarefaction.

The bulk viscosity coefficient of  neutrino-transparent $udse$ matter 
arising from the URCA  processes~\eqref{eq:d_decay}--\eqref{eq:e_capture_s} 
for small-amplitude density oscillations with a frequency $\omega$ is given by 
%------------------------------------------------------
\bea\label{eq:zeta}
\zeta = \frac{C^2}{A}\frac{\gamma}{\omega^2+\gamma^2},
\eea
%------------------------------------------------------ 
which has the classic resonant form depending on two quantities: the
susceptibility prefactor $C^2/A$, which depends only on the equation
of state, and the relaxation rate $\gamma=\lambda_d A$, which depends
additionally on the microscopic interaction rates via the
equilibration coefficient
%------------------------------------------------------ 
\begin{equation}
\lambda\equiv \dfrac{d\Gamma}{d\mu_\Delta},
\end{equation}
%------------------------------------------------------ 
where $\Gamma$ is the total rate of the URCA process.  Analogous
quantities for nonleptonic processes are defined in a similar fashion,
e.g., $\lambda'\equiv {d\Gamma'}/{d\mu'_{\Delta}}$ where $\Gamma'$ is
the rate of nonleptonic process and $\mu'_{\Delta}$ is the deviation
from chemical equilibrium due to nonleptonic processes.  The
susceptibilities are defined as
%------------------------------------------------------
\bea\label{eq:A_def}
A= -\dfrac{1}{n_B}
 \dfrac{\partial \mu_\Delta}{\partial x_u} \bigg\vert_{n_B},\qquad
%\label{eq:C_def}
C = n_B \dfrac{\partial \mu_\Delta}{\partial n_B} \bigg\vert_{x_u},
\eea 
%------------------------------------------------------
where $n_B$ is the baryon density and $x_u$ is the fraction of blue up quarks.
 The derivatives in Eq.~\eqref{eq:A_def} and the relaxation rate $\gamma$ should be computed in $\beta$-equilibrium. (In the regime of interest the density oscillations are supposed to be adiabatic, however, the difference between adiabatic and isothermal susceptibilities is negligible in this case~\cite{Alford2019a,Alford2023}.)

To describe the properties of 2SC quark matter we adopt an 
NJL-type Lagrangian which is given by 
%----------------------------------
\begin{eqnarray}
\label{eq:NJL-Lagrangian}
\mathcal{L}_{Q}&=&\bar\psi(i\gamma^{\mu}\partial_{\mu}-\hat m)\psi 
+G_V(\bar\psi i \gamma^{\mu}\psi)^2\nonumber\\
&+&G_S \sum_{a=0}^8 [(\bar\psi\lambda_a\psi)^2+(\bar\psi i\gamma_5 \lambda_a\psi)^2]\nonumber\\
&+& G_D \sum_{\gamma,c}[\bar\psi_{\alpha}^a i \gamma_5
\epsilon^{\alpha\beta\gamma}\epsilon_{abc}(\psi_C)^b_{\beta}][(\bar\psi_C)^r_{\rho} 
i \gamma_5\epsilon^{\rho\sigma\gamma}\epsilon_{rsc}\psi^8_{\sigma}]\nonumber\\
&-&K \left \{ {\rm det}_{f}[\bar\psi(1+\gamma_5)\psi]+{\rm det}_{f}[\bar\psi(1-\gamma_5)\psi]\right\},
\end{eqnarray}
%----------------------------------
where the quark spinor fields $\psi_{\alpha}^a$ carry color $a = r, g,b$
and flavor ($\alpha= u, d, s$) indices, the matrix of quark current
masses is given by $\hat m= {\rm diag}_f(m_u, m_d, m_s)$, $\lambda_a$
where $ a = 1,..., 8$ are the Gell-Mann matrices in the color space,
and $\lambda_0=(2/3) {\bf 1_f}$. The charge conjugate spinors are
defined as $\psi_C=C\bar\psi^T$ and $\bar\psi_C=\psi^T C$, where
$C=i\gamma^2\gamma^0$ is the charge conjugation matrix.  The original
NJL model, which contains the scalar coupling term with coupling $G_S$
has been extended here to include the 't Hooft interaction with
coupling $K$ and the vector interactions with coupling $G_V$; this
model is referred to as the vector-NJL model. The pairing gap in the
quasiparticle spectrum is $\Delta_c\propto
(\bar\psi_C)_{\alpha}^ai\gamma_5\epsilon^{\alpha\beta
  c}\epsilon_{abc}\psi_{\beta}^b$.  The numerical values of the
parameters of the Lagrangian are the same as in
Ref.~\cite{Bonanno2012} and we fix the strength of the diquark
interaction at $G_D/G_S = 1.1$, this ratio being somewhat larger than
the value 0.75 suggested by the Fierz transformation.  Note that the
larger the gap, the stiffer the equation of state and the larger the
maximum mass of a hybrid star.

We assume 2SC pairing by adopting a diquark-antidiquark interaction
[fourth term in Eq.~\eqref{eq:NJL-Lagrangian}] that has color and
flavor antisymmetric pairing of $u$ and $d$ quarks while $s$ quarks
remain unpaired.

The 2SC pairing gap depends weakly on the density: it varies in the
range $130\le \Delta\le 150$~MeV in the number density range $ 0.6 \le
n_B\le 1.14 $~fm$^{-3}$. It also depends weakly on the temperature in
the range of interest $\Delta \gg T\sim (1-10)$~MeV.  The masses of
quarks in the same density interval vary in the range $10\le
m_{u,d}\le 30$ and $300\le m_s\le 400$~MeV and are weakly dependent on
temperature in the range $1\le T\le 10$~MeV.  Further input
characteristics of the unperturbed quark matter model forming the
background equilibrium are given in Supplemental Material \cite{SM}.
%-------------------------------------------------
\begin{figure}[tbh]  
\begin{center}
\includegraphics[width=0.9\columnwidth,keepaspectratio]{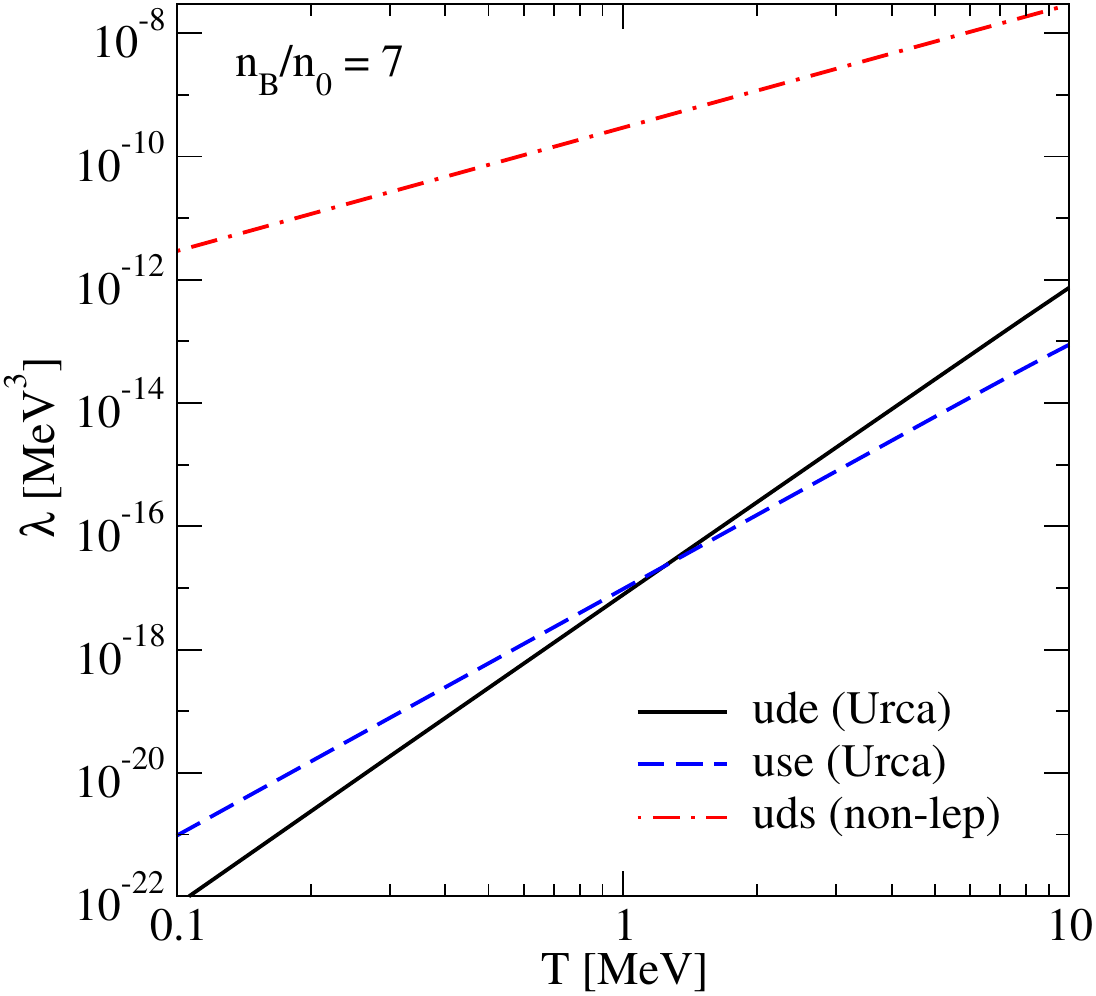}
\caption{ The total equilibration coefficients $\lambda$ for
  processes~\eqref{eq:d_decay} and \eqref{eq:e_capture_d} (solid
  line), for processes~\eqref{eq:s_decay} and \eqref{eq:e_capture_s}
  (dashed line) and for the nonleptonic processes (dash-dotted line)
  as functions of the temperature for baryon density fixed at
  $n_B/n_0=7$. }
\label{fig:lambda} 
\end{center}
\end{figure}

%-------------------------------------------------
\begin{figure}[t]  
\begin{center}
\includegraphics[width=0.9\columnwidth,keepaspectratio]{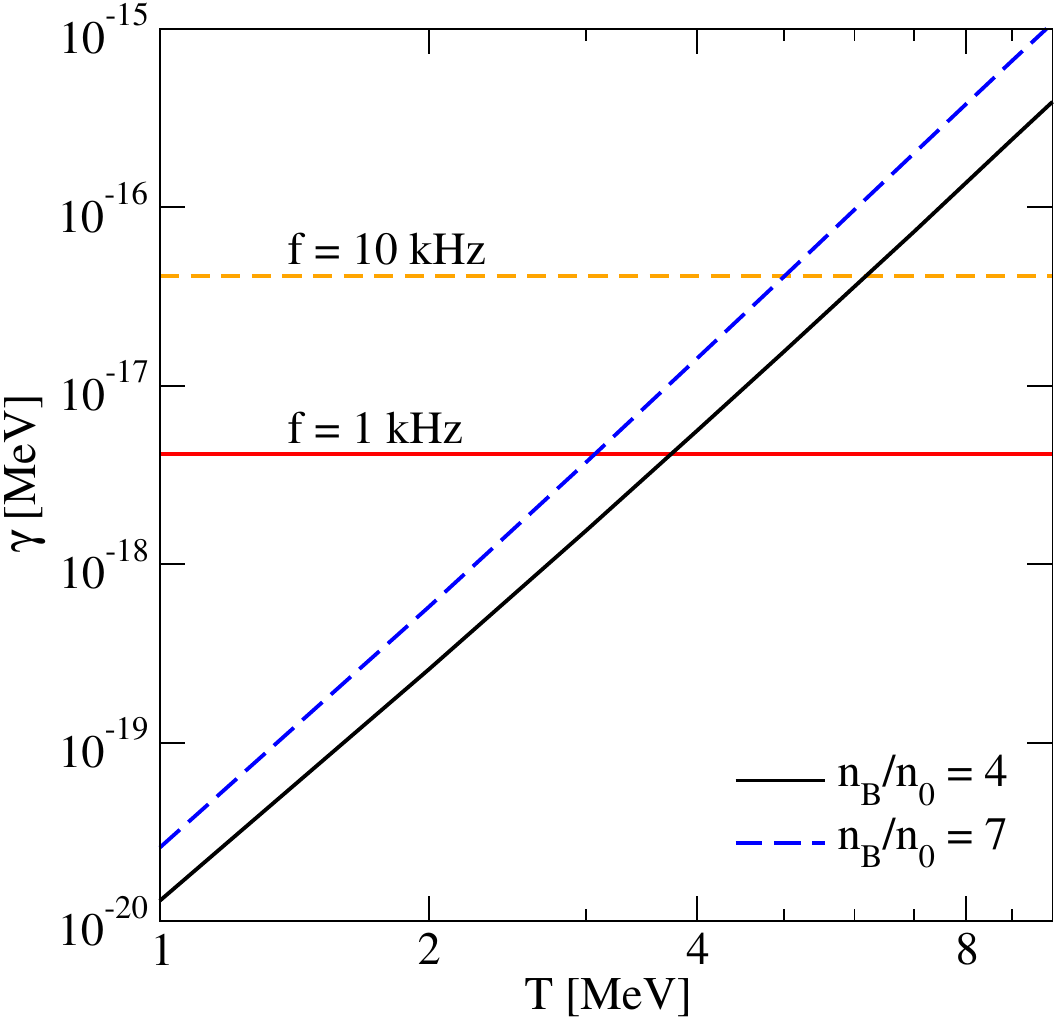}
\caption{ The $\beta$-relaxation rate $\gamma$ as a function of the
  temperature for two values of the baryon density. The horizontal
  lines show where $\gamma=2\pi f$ for selected values of oscillation
  frequency $f= 1$~kHz (solid line) and $f= 10$~kHz (dashed line).}
\label{fig:gamma} 
\end{center}
\end{figure}
%-------------------------------------------------
The equilibration coefficients for the URCA
processes~\eqref{eq:d_decay}--\eqref{eq:e_capture_s} as well as for
the nonleptonic processes~\eqref{eq:non-leptonic} are plotted in
Fig.~\ref{fig:lambda} for fixed baryon density $n_B/n_0=7$. The
nonleptonic processes have a much higher rate than the URCA processes,
as mentioned above, whereas the rates of the $d$-URCA and $s$-URCA
processes are comparable in the temperature range $0.1\leq T\leq
10$~MeV.  This implies that the bulk viscosity produced by the
nonleptonic processes will, for typical BNS merger oscillation
frequencies in the kilohertz  range, be peaked at much lower temperatures
which lie out of the temperature range of interest of BNS mergers.
This argument justifies our assumption that the bulk viscosity at
temperatures $1\leq T \leq 10$~MeV can be computed from URCA processes
only, with the additional constraint $\mu_d=\mu_s$.

To gain physical insight, we quote here approximate analytical
low-temperature results for the equilibration coefficients $\lambda_d
\simeq 0.2\, {G}_F^2\cos^2\!\theta_c \, p_{Fd}^2 T^5$, and $\lambda_s
\simeq 0.03\, {G}_F^2 \sin^2\!\theta_c \,\mu_s^* m_s^{2}T^4$, where
$G_F$ is the Fermi coupling constant and $p_{Ff}$ is the Fermi
momentum of quark of flavor $f=u,d,s$. These formulas were derived
under the assumptions $T\ll \mu_i$, $m_d, m_s\to 0$ and
$\omega_0-\phi_0\ll m_s$ and are accurate to a few percent when
compared to numerical results that do not assume these approximations.
The rates of $d$-URCA processes include an extra power of $T$ because
the scattering matrix element contains the four-product of the momenta
of the $u$-quark and electron, which are nearly collinear at chemical
equilibrium. This near collinearity occurs because the light quarks
are almost massless, with a small angle deviation allowed only by the
thermal smearing of these momenta around the Fermi
spheres. Consequently, the scattering matrix element adds an extra
power of $T$ to the process rates, in addition to the standard $T^5$
scaling for massive particles, which arises from the available
scattering phase space. This difference in temperature scaling leads
to the intersection of the $\lambda_d(T)$ and $\lambda_s(T)$ curves at
a temperature in the MeV range, as seen in Fig.~\ref{fig:lambda}. Two
more aspects are worth noticing: (i) $\lambda_s(T)$ is suppressed by a
factor of $\sin^2\theta_c=0.05$ compared to $\lambda_d(T)$ which has
$\cos^2\theta_c=0.95$ and has a smaller numerical prefactor arising
from the phase-space integration. The remaining factors such as $m_s$,
$p_{Fd}$, and $\mu_s^*$ are all of the same order of magnitude. (ii)
Because the interactions modify the chemical potentials of light
quarks via the isoscalar $\omega$-field, $p_{Fd}=p_{fu}+p_{Fe}$ in
$\beta$-equilibrium for massless quarks, which in turn implies that
the direct URCA channel is only thermally open. In contrast, $s$-URCA
processes are widely open with $p_{Fu}+p_{Fe}-p_{Fs}\geq 70$~MeV as a
result of large $s$-quark mass~\cite{Duncan1983ApJ,Duncan1984ApJ}.

The temperature dependence of the total $\beta$-relaxation rate
$\gamma$ due to the URCA
processes~\eqref{eq:d_decay}--\eqref{eq:e_capture_s} is shown in
Fig.~\ref{fig:gamma}.  It follows a power-law scaling $\gamma\propto
T^{4.5}$ in the temperature range $1\leq T\leq 10$~MeV, and increases
weakly with the density. We see that $\gamma$ matches the constant
angular frequency $\omega=2\pi f$ at a temperature around $T\simeq$3-4~MeV for $f=1\,\kHz$
and around $T\simeq$ 5-6~MeV for $f=10\,\kHz$,
the exact values of the temperature being dependent on the baryon
density.

%-------------------------------------------------
\begin{figure}[t] 
\begin{center}
\includegraphics[width=0.9\columnwidth, keepaspectratio]{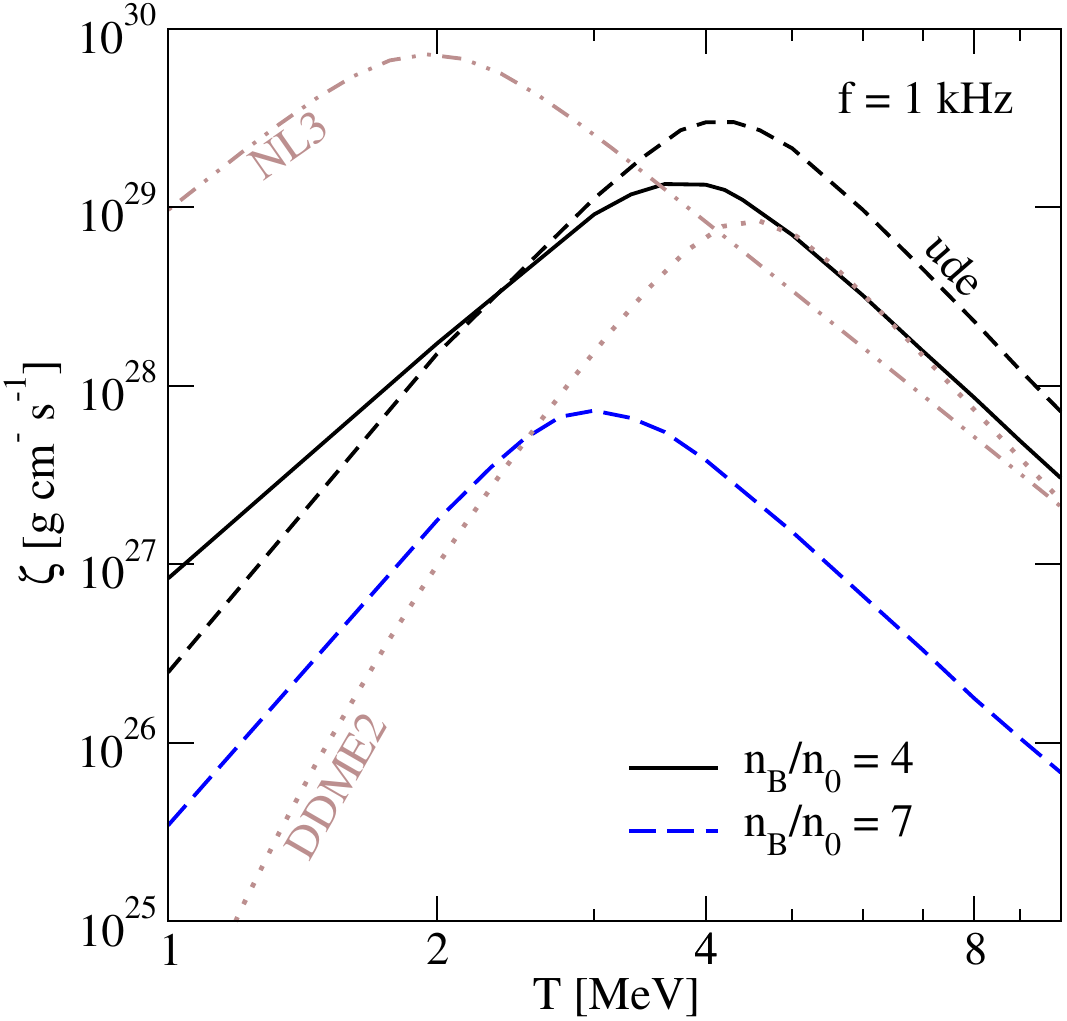}
\caption{The bulk viscosity of $udse$ 2SC quark matter as a function
  of temperature for two values of baryon density. The frequency of
  density oscillations is fixed at $f=1$ kHz.  The bulk viscosity of
  $ude$ matter is shown by short dashed lines at $n_B/n_0=4$.  In
  addition, the bulk viscosity of $npe$ matter from
  Ref.~\cite{Alford2023} is shown for comparison by dotted (model
  DDME2) and dot-dot-dashed (model NL3) lines at density
  $n_B/n_0=4$. }
\label{fig:zeta} 
\end{center}
\end{figure}
%-------------------------------------------------
\begin{figure}[!tbhp] 
\centering
\includegraphics[width=0.9\columnwidth, keepaspectratio]{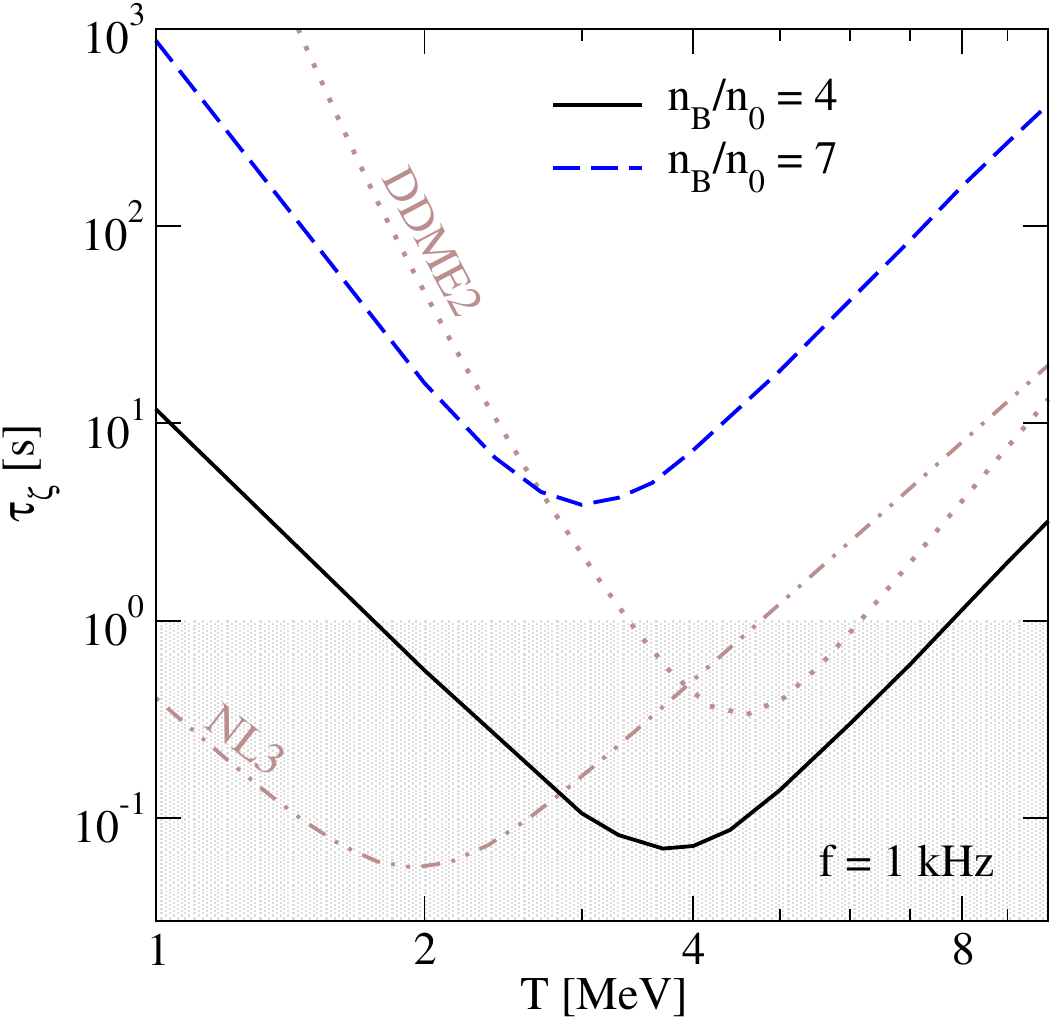}
\caption{ The damping timescale of density oscillations for 2SC quark
  matter as a function of temperature for $f=1$~kHz. The results for
  nucleonic matter from Ref.~\cite{Alford2023} are shown for
  comparison by dotted (model DDME2) and dot-dot-dashed (model
  NL3) lines at density $n_B/n_0=4$. }
\label{fig:tau_damp1} 
%\end{center}
\end{figure}
%-------------------------------------------------
\begin{figure}[!tbhp] 
\centering
\includegraphics[width=0.9\columnwidth, keepaspectratio]{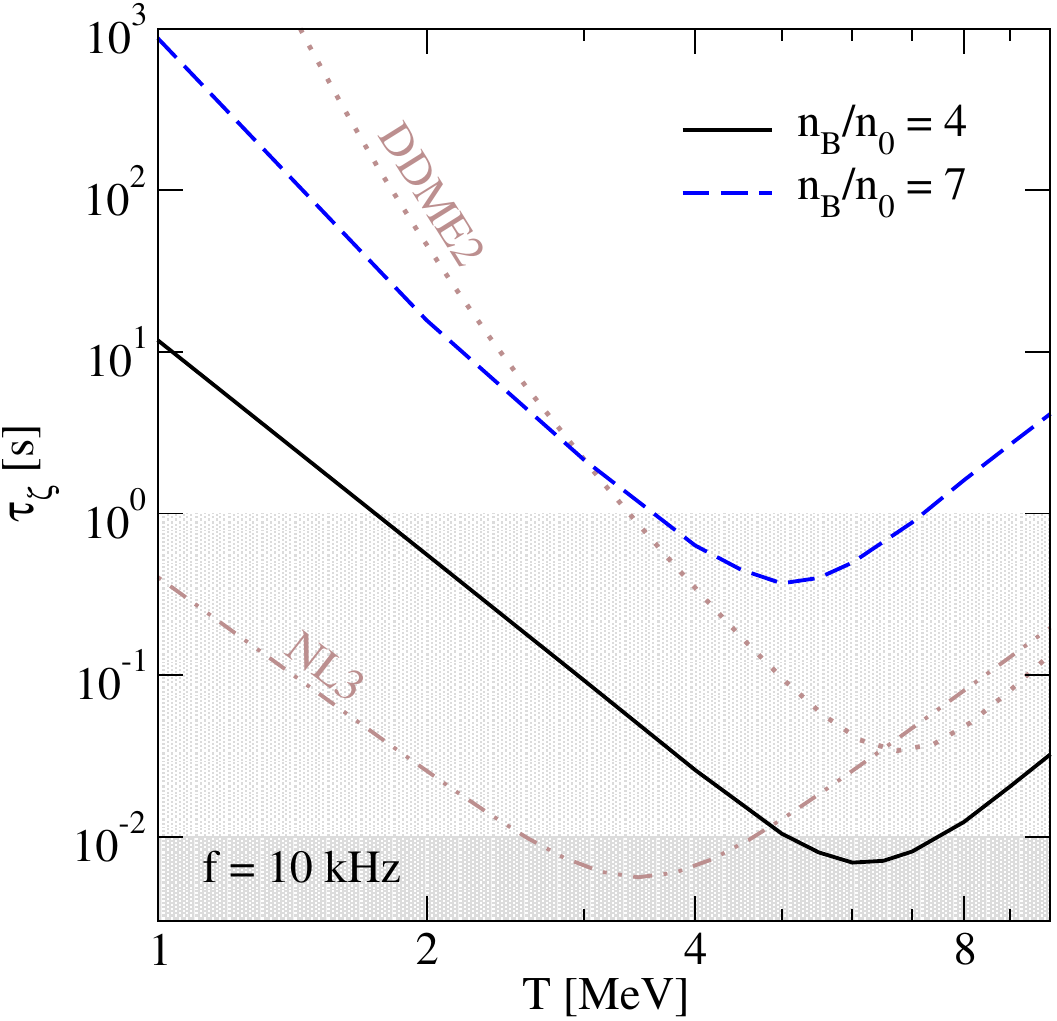}
\caption{ Same as Fig.~\ref{fig:tau_damp1} but for $f=10$~kHz.}
\label{fig:tau_damp10} 
%\end{center}
\end{figure}
%-------------------------------------------------
Figure~\ref{fig:zeta} shows the temperature dependence of the bulk
viscosity of 2SC quark matter for the density oscillation frequency
fixed at a characteristic to BNS mergers value $f=1$~kHz. The bulk
viscosity has a maximum at the temperature where $\omega=\gamma$. The
point of the maximum shifts slowly to lower temperatures with
increasing density. The maximum value and the overall magnitude of the
bulk viscosity is a sensitive function of density through its
dependence on the susceptibility factor $C^2/A$, see
Eq.~\eqref{eq:zeta}.  It decreases with the density, and, e.g., at
$n_B/n_0=7$ the maximum bulk viscosity is more than an order of
magnitude smaller than at $n_B/n_0=4$.  For comparison, we show also
the bulk viscosity of $ude$ matter [\ie, when the
  processes~\eqref{eq:s_decay} and \eqref{eq:e_capture_s} are not
  taken into account] with short dashed lines at $n_B/n_0=4$. We see
that neglecting the $s$-quark contribution underestimates the bulk
viscosity at low temperatures and overestimates it at higher
temperatures. However, the location of the maximum does not change
significantly, because at these temperatures the $d$ URCA and $s$ URCA
processes have comparable rates, and the inclusion of $s$ quarks
increases the relaxation rate by a factor between 1 and 2, leading to
a slight shift of the maximum to lower temperatures.

For comparison, we show also the bulk viscosity of
neutrino-transparent nucleonic matter as was computed in
Ref.~\cite{Alford2023} for models DDME2 (dotted line) and NL3 (dashed
line) for $n_B/n_0=4$. The main difference between these two models is
that the low-temperature direct URCA channel is allowed for NL3 and is
blocked for DDME2 at the density value considered. As a result, the
bulk viscosities for the two models differ significantly at
temperatures $T\leq 4$~MeV, but are rather similar at $T\geq
4$~MeV. At these temperatures, the bulk viscosity of 2SC quark matter
is not very different from the bulk viscosity of nucleonic matter. In
the low-temperature regime, 2SC quark matter's bulk viscosity has
intermediate values compared to the two nucleonic models. The reason
for this is the fact that the $u$-quark fraction is below, but very
close to the direct URCA threshold for the
reactions~\eqref{eq:d_decay} and \eqref{eq:e_capture_d}: the
difference between the Fermi momenta of initial and final state
particles $p_{Fd}-p_{Fu}-p_{Fe}$ is less than 1~MeV for densities
$4\leq n_B/n_0\leq 7$. This implies that already at temperature $T\geq
1$~MeV thermal smearing is sufficient to open up phase space for
light-quark URCA processes.

Since the bulk viscosity comes from semileptonic weak interactions of
the unpaired blue quarks, one would expect that calculations for
unpaired quark matter would see a similar maximum in the bulk
viscosity at MeV temperatures. This is visible in Ref.~\cite{Sad2007b}
(Fig. 3), and also in Ref.~\cite{Hernandez2024} (Fig. 4) though their
maximum bulk viscosity is smaller because their strange quark is
lighter.

Next, we estimate the bulk viscous damping timescales of density
oscillations. The damping timescale is given by
Refs.~\cite{Alford2018a,Alford2019a,Alford2020}
%--------------------------------------------------
\bea\label{eq:damping_time}
\tau_{\zeta} =\frac{1}{9}\frac{Kn_B}{\omega^2\zeta},
\eea 
%--------------------------------------------------
where the incompressibility  is
%--------------------------------------------------
%\bea\label{eq:compress}
$K 
%=9n_B\frac{\partial^2\epsilon}{\partial n_B^2}
=9(\partial P/\partial n_B),
$
%\eea 
%--------------------------------------------------
and $P$ is the pressure.  The damping timescale exhibits a temperature
dependence that is inverse to that of bulk viscosity, as $K$ remains
insensitive to the temperature within the range $1\le T\le 10$~MeV. As
seen from Figs.~\ref{fig:tau_damp1} and \ref{fig:tau_damp10} the
time-scale $\tau_\zeta$ is shortest when the bulk viscosity is
largest, at $T\simeq$ 3-4~MeV for oscillations of frequency
$f=1$~kHz, and at $T\simeq $ 5-6~MeV for $f=10$~kHz. At this minimum,
the bulk viscous damping timescales of 2SC quark matter are in the
range from a few milliseconds to a few hundred milliseconds, which
coincides with the typical short-term evolution timescales of the
postmerger object. The light-shaded region in
Figs.~\ref{fig:tau_damp1} and \ref{fig:tau_damp10} corresponds
$\tau_\zeta\le 1$~s, whereas the heavy-shaded region in
Fig.~\ref{fig:tau_damp10} corresponds to $\tau_\zeta\le 10$~ms.

For a typical oscillation frequency of $f=1$ kHz, our results predict
that bulk viscous damping would be marginally relevant over the short
BNS postmerger evolution timescales of approximately 10 ms, but
noticeable for longer-lived remnants in the temperature range of 2-8
MeV.  For higher frequencies, such as $f=10$ kHz, bulk viscous damping
significantly affects the short-term evolution timescale of
approximately 10 ms within the temperature range of 5-8 MeV at
relatively low densities. It is also noteworthy that the damping
timescales for 2SC quark matter are quite similar to those found for
nucleonic matter.

This implies that, according to our NJL model of quark matter, at
temperatures in the 1-10\,MeV range it will be difficult to
distinguish 2SC quark matter from nuclear matter based on its bulk
viscous dissipation properties.

\section{Conclusions}
\label{sec:conclusions}

In this work, we studied the bulk viscosity of neutrino-transparent 2SC quark matter driven by direct URCA processes in the temperature range $1\leq T\leq 10$MeV and density range $4n_0\leq n_B\leq 7n_0$. This complements our recent study of relativistic $npe\mu$ matter~\cite{Alford2023}. Here, we focused on direct URCA rates, including the strange sector and the effects of color superconductivity in the red/green up and down quarks. Blue quarks primarily drive the bulk viscosity, although gapless superconductivity may allow contributions from red-green pairs~\cite{Jaikumar2006}.

We found that bulk viscosity in $udse$ matter peaks at $T=3{-}6$~MeV, where URCA process rates resonate with the characteristic frequencies of density oscillations in BNS postmerger objects. The resulting damping timescales, down to tens of milliseconds, suggest that bulk viscosity may significantly contribute to energy dissipation in BNS mergers at these temperatures.

The bulk viscosity and damping time for 2SC quark matter closely
resemble those of nucleonic matter \cite{Alford2023}, making it
challenging to distinguish these states via their bulk viscous
behavior. However, at lower temperatures, quark matter shows a second
resonant peak in bulk viscosity at $T\sim 0.05$\, MeV, absent in
nuclear matter \cite{Alford:2006gy,Rojas2024,Hernandez2024}.

Interestingly, semileptonic rates for unpaired quark matter may be
similar to those for 2SC matter, suggesting that unpaired quark matter
at $T\approx 1$ to $10$~MeV may also be difficult to distinguish from
nuclear matter. A recent study of unpaired quark
matter~\cite{Hernandez2024} found a peak in bulk viscosity in this
temperature range, though with longer damping times, likely due to
different susceptibilities in their model. Future studies should
address these model-dependent uncertainties in quark matter
phenomenology under BNS merger conditions.

\section*{Acknowledgments}

M.~A. is partly supported by the U.S. Department of Energy, Office of
Science, Office of Nuclear Physics under Award No. DE-FG02-05ER41375.
A.~H. and A.~S. were supported by the Volkswagen Foundation (Hannover,
Germany) Grant No.  96 839.  A. S. and S.~T.  acknowledge the support
of the Polish National Science Centre (NCN) Grant
2020/\-37/B/ST9/01937.  A.~S. acknowledges Grant No. SE 1836/5-3 from
Deutsche For\-schungs\-gemeinschaft (DFG).  S.~T. is supported by the
IMPRS for Quantum Dynamics and Control at Max Planck Institute for the
Physics of Complex Systems, Dresden, Germany.

\providecommand{\href}[2]{#2}\begingroup\raggedright\endgroup

\newpage
\begin{widetext}
\section*{\large Supplemental Material\\
Bulk viscosity of two-color superconducting quark matter in neutron star mergers}

Here, we show 
 the particle fractions of unpaired particles and their chemical potentials in finite-temperature 
$\beta$-equilibrated 2SC quark matter~\cite{Alford2007} within the vector-interaction-improved NJL model described in detail in Ref.~\cite{Bonanno2012}.
These are the key input characteristics of the unperturbed quark matter model needed for analyzing perturbations.

The quark (thermodynamic) chemical potentials  are given by
%----------------------------------
\bea
\mu_{f,c}=\frac{1}{3}\mu_B +\mu_Q Q_f+\mu_3T^c_3+\mu_8T^c_8,
\eea
%----------------------------------
with $\mu_B$ and $\mu_Q$ being the baryon and charge chemical potential and
%----------------------------------
\bea
&&Q_f=\operatorname{diag}_f\left(\frac{2}{3},-\frac{1}{3},-\frac{1}{3}\right), 
\\ 
&&T_3^c=\frac{1}{2} \operatorname{diag}_c(1,-1,0),\,\, T_8^c=\frac{1}{2 \sqrt{3}} \operatorname{diag}_c(1,1,-2).
\eea
%----------------------------------
The quark effective chemical potentials are modified by the vector fields ${\mu}^*=$ $\operatorname{diag}_f\left(\mu_u-\omega_0, \mu_d-\omega_0, \mu_s-\phi_0\right)$. 

%-------------------------------------------------
\begin{figure}[bht] 
\begin{center}
\includegraphics[width=0.45\columnwidth, keepaspectratio]{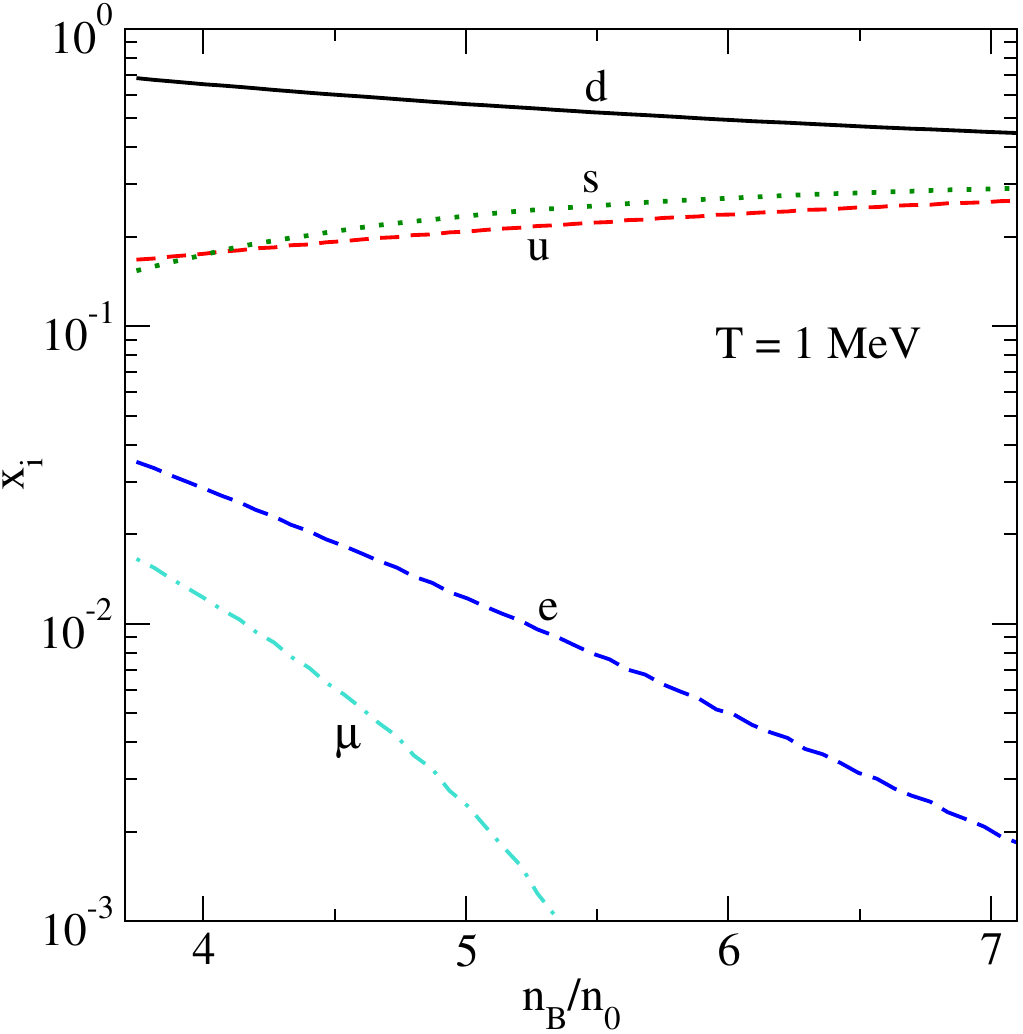}
\caption{Particle fractions of unpaired particles in finite-temperature 
$\beta$-equilibrated 2SC quark matter as functions of the baryon density 
$n_B$ (in units of nuclear saturation density $n_0=0.16$~fm$^{-3}$) at $T=1$~MeV. 
They are insensitive to the value of the temperature within the range $1\le T\le 10$~MeV.}
\label{fig:fractions} 
\end{center}
\end{figure}

Figure~\ref{fig:fractions} shows the finite-temperature composition of 
the 2SC phase under $\beta$-equilibrium, relevant for the computation 
of the bulk viscosity, where we show only the population of the blue 
(unpaired color) $u$ and $d$ quarks, as well as the population 
of $s$ quarks of blue color.  (The blue $s$-quark density differs slightly from  
red-green $s$-quark densities, with the difference 
in chemical potentials being on the order of 2\%, 
approximately 10 MeV.)
Note that the substantial $s$-quark population is present in matter which 
quickly equilibrates with the light flavor sector via the non-leptonic processes.  It is also seen that $s$-quarks reduce the lepton fraction in matter,  as already noticed long ago~\cite{Duncan1983ApJ, Duncan1984ApJ}.

Figure~\ref{fig:chemical-potentials} displays the effective chemical potentials of the species in $\beta$-equilibrated 2SC matter. We show only the chemical potentials of blue quarks, which are relevant for the computation of the bulk viscosity.  The temperature dependence of the chemical potentials is weak in the shown density range. Note that the fast equilibration via non-leptonic processes implies $\mu_d=\mu_s$ and the difference in the effective chemical potentials of $d$ and $s$ quarks seen in Fig.~\ref{fig:chemical-potentials} is 
seen in Fig.~\ref{fig:chemical-potentials} is due to their coupling to $\omega$ and $\phi$ mesons.

%-------------------------------------------------
\begin{figure}[bht] 
\begin{center}
\includegraphics[width=0.45\columnwidth, keepaspectratio]{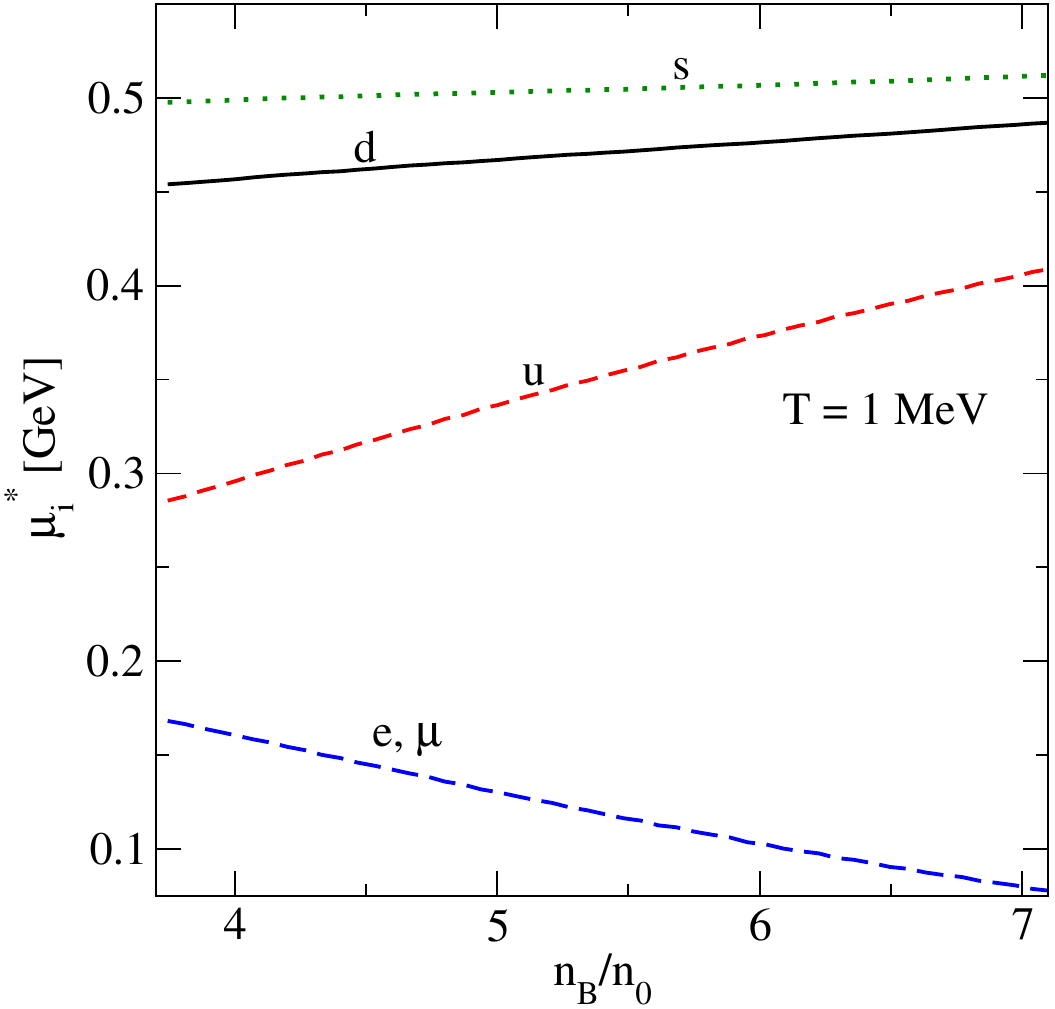}
\caption{The same as in Fig.~\ref{fig:fractions} but 
for effective chemical potentials of unpaired particles. They depend weakly on the temperature in the range $1\le T\le 10$~MeV .}
\label{fig:chemical-potentials} 
\end{center}   
\end{figure}

\end{widetext}


\begin{thebibliography}{10}

\bibitem{Abbott2017}
{The LIGO Scientific Collaboration} and {The Virgo Collaboration},
  \emph{GW170817: Observation of gravitational waves from a binary neutron star
  inspiral},
  \href{http://dx.doi.org/10.1103/PhysRevLett.119.161101}{\emph{Phys. Rev.
  Lett.} {\bf 119} (2017) 161101}.

\bibitem{Faber2012}
J.~A. Faber and F.~A. Rasio, \emph{{Binary Neutron Star Mergers}},
  \href{http://dx.doi.org/10.12942/lrr-2012-8}{\emph{Living Rev. Rel.} {\bf 15}
  (2012) 8}, [\href{http://arxiv.org/abs/1204.3858}{{\tt 1204.3858}}].

\bibitem{Baiotti2017}
L.~Baiotti and L.~Rezzolla, \emph{{Binary neutron-star mergers: a review of
  Einstein's richest laboratory}},
  \href{http://dx.doi.org/10.1088/1361-6633/aa67bb}{\emph{Rept. Prog. Phys.}
  {\bf 80} (2017) 096901}, [\href{http://arxiv.org/abs/1607.03540}{{\tt
  1607.03540}}].

\bibitem{Baiotti2019}
L.~{Baiotti}, \emph{{Gravitational waves from neutron star mergers and their
  relation to the nuclear equation of state}},
  \href{http://dx.doi.org/10.1016/j.ppnp.2019.103714}{\emph{Progress in
  Particle and Nuclear Physics} {\bf 109} (2019) 103714},
  [\href{http://arxiv.org/abs/1907.08534}{{\tt 1907.08534}}].

\bibitem{Bauswein2019}
A.~{Bauswein}, N.-U.~F. {Bastian}, D.~B. {Blaschke}, K.~{Chatziioannou}, J.~A.
  {Clark}, T.~{Fischer} et~al., \emph{{Identifying a First-Order Phase
  Transition in Neutron-Star Mergers through Gravitational Waves}},
  \href{http://dx.doi.org/10.1103/PhysRevLett.122.061102}{\emph{\prl} {\bf 122}
  (2019) 061102}, [\href{http://arxiv.org/abs/1809.01116}{{\tt 1809.01116}}].

\bibitem{Radice2023}
D.~Radice and S.~Bernuzzi, \emph{{Ab-initio General-relativistic
  Neutrino-radiation Hydrodynamics Simulations of Long-lived Neutron Star
  Merger Remnants to Neutrino Cooling Timescales}},
  \href{http://dx.doi.org/10.3847/1538-4357/ad0235}{\emph{Astrophys. J.} {\bf
  959} (2023) 46}, [\href{http://arxiv.org/abs/2306.13709}{{\tt 2306.13709}}].

\bibitem{Radice2024}
D.~Radice and I.~Hawke, \emph{{Turbulence modelling in neutron star merger
  simulations}}, \href{http://dx.doi.org/10.1007/s41115-023-00019-9}{\emph{Liv.
  Rev. Comput. Astrophys.} {\bf 10} (2024) 1},
  [\href{http://arxiv.org/abs/2402.03224}{{\tt 2402.03224}}].

\bibitem{Alford2018a}
M.~G. {Alford}, L.~{Bovard}, M.~{Hanauske}, L.~{Rezzolla} and K.~{Schwenzer},
  \emph{{Viscous Dissipation and Heat Conduction in Binary Neutron-Star
  Mergers}},
  \href{http://dx.doi.org/10.1103/PhysRevLett.120.041101}{\emph{\prl} {\bf 120}
  (2018) 041101}, [\href{http://arxiv.org/abs/1707.09475}{{\tt 1707.09475}}].

\bibitem{Alford2020}
M.~Alford, A.~Harutyunyan and A.~Sedrakian, \emph{{Bulk Viscous Damping of
  Density Oscillations in Neutron Star Mergers}},
  \href{http://dx.doi.org/10.3390/particles3020034}{\emph{Particles} {\bf 3}
  (2020) 500--517}, [\href{http://arxiv.org/abs/2006.07975}{{\tt 2006.07975}}].

\bibitem{Alford2021a}
M.~G. {Alford} and A.~{Haber}, \emph{{Strangeness-changing rates and hyperonic
  bulk viscosity in neutron star mergers}},
  \href{http://dx.doi.org/10.1103/PhysRevC.103.045810}{\emph{\prc} {\bf 103}
  (2021) 045810}, [\href{http://arxiv.org/abs/2009.05181}{{\tt 2009.05181}}].

\bibitem{Alford2021c}
M.~{Alford}, A.~{Harutyunyan} and A.~{Sedrakian}, \emph{{Bulk viscosity from
  URCA processes: npe{\ensuremath{\mu}} matter in the neutrino-trapped
  regime}}, \href{http://dx.doi.org/10.1103/PhysRevD.104.103027}{\emph{\prd}
  {\bf 104} (2021) 103027}, [\href{http://arxiv.org/abs/2108.07523}{{\tt
  2108.07523}}].

\bibitem{Alford2023}
M.~{Alford}, A.~{Harutyunyan} and A.~{Sedrakian}, \emph{{Bulk viscosity from
  URCA processes: n p e {\ensuremath{\mu}} matter in the neutrino-transparent
  regime}}, \href{http://dx.doi.org/10.1103/PhysRevD.108.083019}{\emph{\prd}
  {\bf 108} (Oct., 2023) 083019}, [\href{http://arxiv.org/abs/2306.13591}{{\tt
  2306.13591}}].

\bibitem{Most2022}
E.~R. {Most}, S.~P. {Harris}, C.~{Plumberg}, M.~.~G. {Alford}, J.~{Noronha},
  J.~{Noronha-Hostler} et~al., \emph{{Projecting the likely importance of
  weak-interaction-driven bulk viscosity in neutron s\ tar mergers}},
  \href{http://dx.doi.org/10.1093/mnras/stab2793}{\emph{\mnras} {\bf 509}
  (Jan., 2022) 1096--1108}, [\href{http://arxiv.org/abs/2107.05094}{{\tt
  2107.05094}}].

\bibitem{Celora2022}
T.~{Celora}, I.~{Hawke}, P.~C. {Hammond}, N.~{Andersson} and G.~L. {Comer},
  \emph{{Formulating bulk viscosity for neutron star simulations}},
  \href{http://dx.doi.org/10.1103/PhysRevD.105.103016}{\emph{\prd} {\bf 105}
  (May, 2022) 103016}, [\href{http://arxiv.org/abs/2202.01576}{{\tt
  2202.01576}}].

\bibitem{Chabanov2023}
M.~{Chabanov} and L.~{Rezzolla}, \emph{{Impact of bulk viscosity on the
  postmerger gravitational-wave signal from merging neutron stars}},
  \href{http://dx.doi.org/10.48550/arXiv.2307.10464}{\emph{arXiv e-prints}
  (July, 2023) arXiv:2307.10464}, [\href{http://arxiv.org/abs/2307.10464}{{\tt
  2307.10464}}].

\bibitem{Sawyer1989}
R.~F. {Sawyer}, \emph{{Bulk viscosity of hot neutron-star matter and the
  maximum rotation rates of neutron stars}},
  \href{http://dx.doi.org/10.1103/PhysRevD.39.3804}{\emph{\prd} {\bf 39} (1989)
  3804--3806}.

\bibitem{Madsen1992}
J.~{Madsen}, \emph{{Bulk viscosity of strange quark matter, damping of quark
  star vibration, and the maximum rotation rate of pulsars}},
  \href{http://dx.doi.org/10.1103/PhysRevD.46.3290}{\emph{\prd} {\bf 46} (Oct.,
  1992) 3290--3295}.

\bibitem{Drago2005}
A.~{Drago}, A.~{Lavagno} and G.~{Pagliara}, \emph{{Bulk viscosity in hybrid
  stars}}, \href{http://dx.doi.org/10.1103/PhysRevD.71.103004}{\emph{\prd} {\bf
  71} (May, 2005) 103004}, [\href{http://arxiv.org/abs/astro-ph/0312009}{{\tt
  astro-ph/0312009}}].

\bibitem{Alford:2006gy}
M.~G. Alford and A.~Schmitt, \emph{{Bulk viscosity in 2SC quark matter}},
  \href{http://dx.doi.org/10.1088/0954-3899/34/1/005}{\emph{J. Phys.} {\bf G34}
  (2007) 67--102}, [\href{http://arxiv.org/abs/nucl-th/0608019}{{\tt
  nucl-th/0608019}}].

\bibitem{Blaschke2007}
D.~B. {Blaschke} and J.~{Berdermann}, \emph{{Neutrino emissivity and bulk
  viscosity of iso-CSL quark matter in neutron stars}},  in \emph{QCDatWORK
  2007: International Workshop on Quantum Chromodynamics: Theory and
  Experiment} (P.~{Colangelo}, D.~{Creanza}, F.~{de Fazio}, R.~A. {Fini} and
  E.~{Nappi}, eds.), vol.~964 of \emph{American Institute of Physics Conference
  Series}, pp.~290--295, AIP, Nov., 2007.
\newblock \href{http://arxiv.org/abs/0710.5243}{{\tt 0710.5243}}.
\newblock \href{http://dx.doi.org/10.1063/1.2823866}{DOI}.

\bibitem{Sad2007a}
B.~A. {Sa'd}, I.~A. {Shovkovy} and D.~H. {Rischke}, \emph{{Bulk viscosity of
  spin-one color superconductors with two quark flavors}},
  \href{http://dx.doi.org/10.1103/PhysRevD.75.065016}{\emph{\prd} {\bf 75}
  (Mar., 2007) 065016}, [\href{http://arxiv.org/abs/astro-ph/0607643}{{\tt
  astro-ph/0607643}}].

\bibitem{Sad2007b}
B.~A. {Sa'd}, I.~A. {Shovkovy} and D.~H. {Rischke}, \emph{{Bulk viscosity of
  strange quark matter: URCA versus nonleptonic processes}},
  \href{http://dx.doi.org/10.1103/PhysRevD.75.125004}{\emph{\prd} {\bf 75}
  (June, 2007) 125004}, [\href{http://arxiv.org/abs/astro-ph/0703016}{{\tt
  astro-ph/0703016}}].

\bibitem{Huang2010}
X.-G. {Huang}, M.~{Huang}, D.~H. {Rischke} and A.~{Sedrakian},
  \emph{{Anisotropic hydrodynamics, bulk viscosities, and r-modes of strange
  quark stars with strong magnetic fields}},
  \href{http://dx.doi.org/10.1103/PhysRevD.81.045015}{\emph{\prd} {\bf 81}
  (Feb., 2010) 045015}, [\href{http://arxiv.org/abs/0910.3633}{{\tt
  0910.3633}}].

\bibitem{Wang2010a}
X.~{Wang}, H.~{Malekzadeh} and I.~A. {Shovkovy}, \emph{{Nonleptonic weak
  processes in spin-one color superconducting quark matter}},
  \href{http://dx.doi.org/10.1103/PhysRevD.81.045021}{\emph{\prd} {\bf 81}
  (Feb., 2010) 045021}, [\href{http://arxiv.org/abs/0912.3851}{{\tt
  0912.3851}}].

\bibitem{Wang2010b}
X.~{Wang} and I.~A. {Shovkovy}, \emph{{Bulk viscosity of spin-one color
  superconducting strange quark matter}},
  \href{http://dx.doi.org/10.1103/PhysRevD.82.085007}{\emph{\prd} {\bf 82}  (Oct., 2010) 085007}, [\href{http://arxiv.org/abs/1006.1293}{{\tt
  1006.1293}}].

\bibitem{Rojas2024}
J.~{Cruz Rojas}, T.~{Gorda}, C.~{Hoyos}, N.~{Jokela}, M.~{J{\"a}rvinen}, et~al., \emph{{Estimate for the bulk viscosity of strongly coupled quark matter}},
  \href{http://dx.doi.org/10.1103/PhysRevLett.133.071901}{\emph{\prl} {\bf 133}  (Aug, 2024) 071901},
  [\href{http://arxiv.org/abs/2402.00621}{{\tt  2402.00621}}].


\bibitem{Hernandez2024}
J.~L. {Hern{\'a}ndez}, C.~{Manuel} and L.~{Tolos}, \emph{{Damping of density
  oscillations from bulk viscosity in quark matter}},
  \href{http://dx.doi.org/10.1103/PhysRevD.109.123022}{\emph{\prd} {\bf 109}
  (June, 2024) 123022}, [\href{http://arxiv.org/abs/2402.06595}{{\tt
  2402.06595}}].

\bibitem{Alford2007}
M.~G. Alford, A.~Schmitt, K.~Rajagopal and T.~Sch\"afer, \emph{{Color
  superconductivity in dense quark matter}},
  \href{http://dx.doi.org/10.1103/RevModPhys.80.1455}{\emph{Rev. Mod. Phys.}
  {\bf 80} (2008) 1455--1515}, [\href{http://arxiv.org/abs/0709.4635}{{\tt
  0709.4635}}].

\bibitem{Bonanno2012}
L.~Bonanno and A.~Sedrakian, \emph{Composition and stability of hybrid stars
  with hyperons and quark color-superconductivity}, \href{		https://doi.org/10.1051/0004-6361/201117832}{\emph{A\&A} {\bf 539}
  (2012) A16}.

  \bibitem{SM}
See Supplemental Material at [http://link.aps.org/ supplemental/10.1103/xxxx] for the particle fractions of unpaired particles and their chemical potential in finite-temperature $\beta$-equilibrated 2SC quark matter, which are
used for analyzing perturbations from an equilibrium background.

\bibitem{Duncan1983ApJ}
R.~C. {Duncan}, S.~L. {Shapiro} and I.~{Wasserman}, \emph{{Equilibrium
  composition and neutrino emissivity of interacting quark matter in neutron
  stars}}, \href{http://dx.doi.org/10.1086/160875}{\emph{\apj} {\bf 267} (Apr.,
  1983) 358--370}.

\bibitem{Duncan1984ApJ}
R.~C. {Duncan}, I.~{Wasserman} and S.~L. {Shapiro}, \emph{{Neutrino emissivity
  of interacting quark matter in neutron stars. II - Finite neutrino momentum
  effects}}, \href{http://dx.doi.org/10.1086/161850}{\emph{\apj} {\bf 278}
  (Mar., 1984) 806--812}.

\bibitem{Alford2019a}
M.~G. {Alford} and S.~P. {Harris}, \emph{{Damping of density oscillations in
  neutrino-transparent nuclear matter}},
  \href{http://dx.doi.org/10.1103/PhysRevC.100.035803}{\emph{\prc} {\bf 100}
  (2019) 035803}, [\href{http://arxiv.org/abs/1907.03795}{{\tt 1907.03795}}].



\bibitem{Jaikumar2006}
P.~{Jaikumar}, C.~D. {Roberts} and A.~{Sedrakian}, \emph{{Direct URCA neutrino
  rate in color superconducting quark matter}},
  \href{http://dx.doi.org/10.1103/PhysRevC.73.042801}{\emph{\prc} {\bf 73}
  (Apr., 2006) 042801}, [\href{http://arxiv.org/abs/nucl-th/0509093}{{\tt
  nucl-th/0509093}}].

\end{thebibliography}
\end{document}